\documentclass{article}

\usepackage{PRIMEarxiv}

\usepackage[utf8]{inputenc} 
\usepackage[T1]{fontenc}    
\usepackage[hidelinks]{hyperref}       
\usepackage{url}            
\usepackage{booktabs}       
\usepackage{amsfonts}       
\usepackage{nicefrac}       
\usepackage{microtype}      
\usepackage{lipsum}
\usepackage{fancyhdr}       
\usepackage{graphicx}       
\graphicspath{{media/}}     
\usepackage{amssymb}
\usepackage{mathtools}      
\usepackage{cleveref}           
\usepackage{caption}
\usepackage{subcaption}
\usepackage{multirow}
\usepackage{float}
\usepackage[dvipsnames]{xcolor}
\usepackage{empheq}
\usepackage{soul}
\usepackage{amsmath}

\usepackage{chngcntr}

\usepackage{natbib}
\setcitestyle{authoryear,open={(},close={)}}

\begin{document}
\title{Roughness evolution induced by third-body wear}

\newcommand{\blind}{0}
		\if0\blind
		{
			\author{
			{\Large
			Joaquin Garcia-Suarez, Tobias Brink\thanks{Present address: Max-Planck-Institut f\"ur Eisenforschung GmbH, Max-Planck-Stra\ss{}e 1, D-40237 D\"usseldorf, Germany}   ,
			Jean-François Molinari 
			}\\
			Institute of Civil Engineering, Institute of Materials,\\
            \'{E}cole Polytechnique F\'{e}d\'{e}rale de Lausanne (EPFL), CH 1015 Lausanne, 
            Switzerland 
            }
			\date{}
			\maketitle
		} \fi

\begin{abstract}
Surface roughness is a key factor when it comes to friction and wear, as well as to other physical properties. 
These phenomena are controlled by mechanisms acting at small scales, in which the topography of apparently-flat surfaces is revealed. 
Roughness in natural surfaces has been reported to conform to self-affine statistics in a wide variety of settings (ranging from earthquake physics to micro-electro-mechanical devices), meaning that the height profile can be described using a spectrum where the amplitude is proportional to its wavelength raised to a constant power, which is related to a statistical parameter named Hurst exponent.
We analyze the roughness evolution in atomistic surfaces during molecular dynamics simulations of wear. 
%
Both pairs of initially-flat and initially-rough surfaces in contact are worn by a third body formed by particles trapped between them during relative sliding. 
During the first sliding stages, the particles trapped between the first bodies scratch the surfaces. 
Once the former become coated with atoms from the latter, the wear process slows down and becomes ``adhesive-like''.
The initial particle sizes are consistent with the minimum size to be expected for the debris, but tend to grow by material removal from the surfaces and to agglomerate.
We show that, for the particular configurations under consideration, the surface roughness seems to converge to a steady state characterized by Hurst exponent close to 0.8, independently of the initial conditions.
%
\end{abstract}


\section{Introduction}



Understanding the geometry and evolution of rough surfaces is an active research endeavor in tribology \citep{Persson:2005,Bonamy:2011,Renard:2013,Candela:2016,Aghababaei:2023} as roughness mediates friction, wear and lubrication \citep{Godet:1984}.  
Since the pioneering work of Bowden and Tabor, the response of sliding surfaces is known to depend on the real contact area between the surfaces \citep{Tabor&Bowden}, which, due to their roughness, is but a small percentage of the apparent contact area \citep{Dieterich-Kilgore}. 
In natural surfaces \citep{Renard:2013,Candela:2016},  across a wide range of scales, roughness appears to follow a fractal distribution (we refer in this case to ``self-affine surfaces'') which can be characterized in terms of a power law of the wavelengths whose exponent relates to the so-called ``Hurst exponent'', see \cite{mandelbrot:1968}. 
A number of factors have been put forward to explain how these particular statistics arise in nature: material heterogeneity \citep{Sundaram:2012}, plastic mechanisms \citep{Irani:2019,Hinkle:2020}, fracture \citep{Bouchaud:1994}, and corrosion \citep{Toloei:2013}, among others. 
\cite{Hinkle:2020} proved that fractal roughness can arise from inelastic deformation induced by simple compression, combined with material discreteness and heterogeneity. 

Recently, molecular dynamic (MD) models with simplified potentials have enabled \textit{in silico} experiments in which the transition from wear regimes (from asperity plastic smoothing to fracture-induced debris creation) can be observed \citep{critical_length_scale,PNAS,Breakdown}. In this context, roughness evolution of 1D surfaces was studied \citep{Enrico,Milanese:2020}, reporting the generation of self-affine surfaces starting from contacting asperities that are subsequently sheared during surface relative sliding, which gives rise to a third body and whose rolling ends up wearing the surfaces through a process of tearing of shallow clumps of atoms \citep{Milanese:2020b}, leading eventually to the attainment of a steady-state roughness with fractal characteristics.

\begin{figure}
    \centering
    \includegraphics[width=\linewidth]{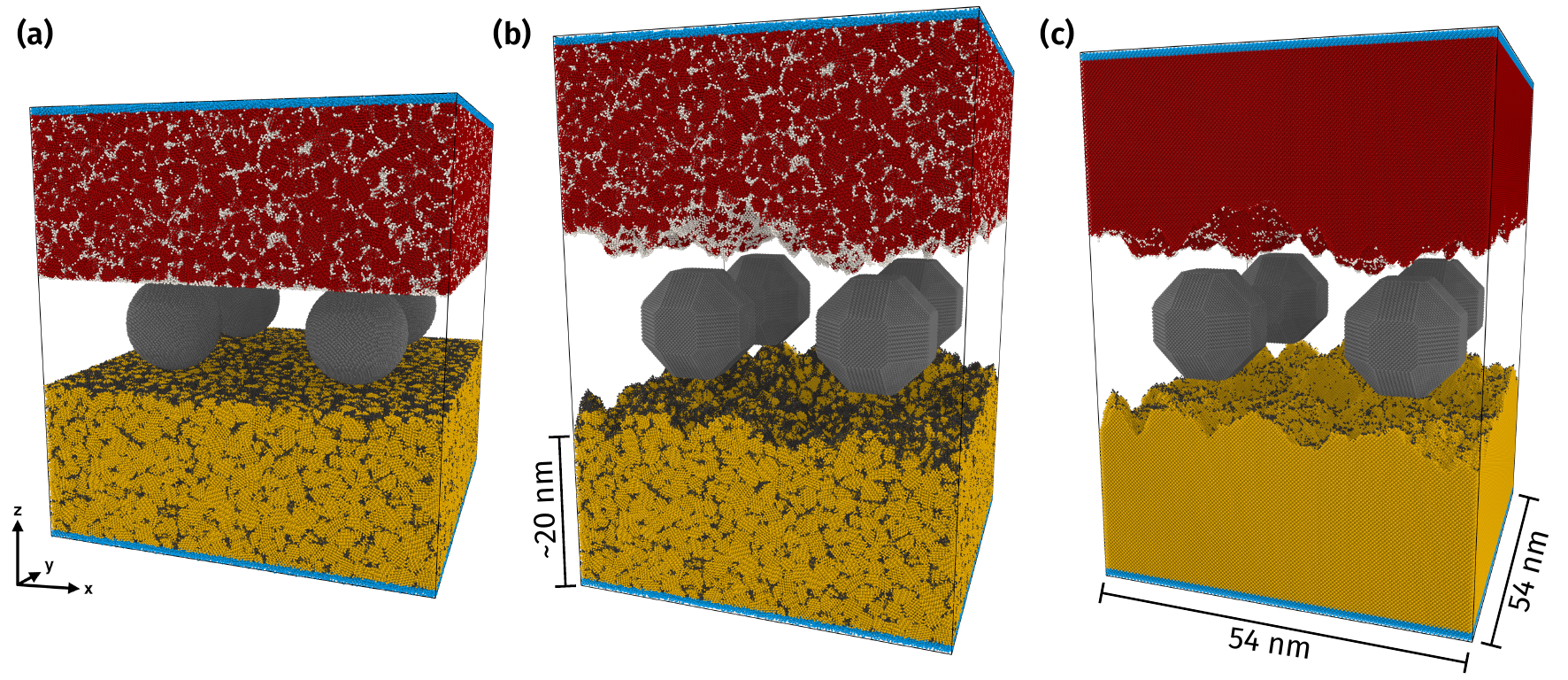}
    \caption{Setup of the sliding contact simulations. In all cases, two blocks of bulk material (yellow: bottom block, red: top block) were put in contact with rigid wear particles (gray). Blue areas indicate the boundaries where force and displacement were imposed. We used different initial setups: (a) Nanocrystalline first bodies (grain boundaries indicated by black and gray atoms) starting from a flat surface and round wear particles, (b) nanocrystalline first bodies starting from artificial surface roughness and polyhedral wear particles, and (c) the same as (b) but with single crystalline first bodies.}
    \label{fig:setup}
\end{figure}

\cite{Milanese:2020} was solely concerned with 2D geometries (1D roughness). Recently, \cite{Brink:2022} performed 3D simulations, starting from a configuration with two surfaces in relative sliding motion with pre-formed third bodies in between (\Cref{fig:setup}). 
This work considered both surfaces that were initially rough and flat as well as third-body particles simultaneously, hence a setting that departs starkly from idealized configurations \citep{Sorensen:1996,Mo:2009,Stoyanov:2013,Eder:2015,Yang:2016,Robbins:2017,Aghababaei:2019}. 
Among the many insights provided by these simulations, the appearance of cylindrical rolling particles as those observed in experiments \citep{Zanoria:1993,Zanoria_a:1995,Zanoria_b:1995} and the attainment of an apparent steady-state roughness regime stood out. 
%
The steady-state is reached after substantial material transfer from the surface to the debris particles. 
This also translates into decimation of the larger topographical features (those associated to the wavelength of the order of the diameter of the debris); 
this also implies an apparent ``flattening'' of the surface that, unlike previous results \citep{Sorensen:1996,Spijker:2011,Stoyanov:2013}, is not only associated to atomistic mechanisms but also to debris creation and its coating by surface atoms upon subsequent sliding.  


Even though the original paper \citep{Brink:2022} was focused on quantifying frictional forces and surface wear, the surface topography state was extracted at regular increments of the simulations (see Methods section). 
This manuscript presents the post-processing of the roughness and characterizes the steady-state regimes quantitatively. 
\Cref{Sec:methods} briefly reviews the computations' specificities and presents the topography spectral analysis techniques to be utilized. The surface changes as well as the corresponding Hurst exponent evolution are reported in \Cref{Sec:results} and discussed in \Cref{Sec:discussion}. Conclusions and future work directions are presented in \Cref{Sec:final_remarks}.

\section{Methods}
\label{Sec:methods}

\subsection{Data generation}
\label{Sec:methods_tobias}

Rough surfaces were generated by MD simulations of rigid particles rolling on surfaces made out of a silicon-like model material \citep{Brink:2022}. We used LAMMPS \citep{Thompson2022} with a modified \citep{Holland1998, Holland1998erratum} Stillinger--Weber \citep{Si_potential} potential using GPU acceleration \citep{Brown2011, Brown2013}. While the modified potential does not reproduce the properties of silicon, it can model the brittle fracture (cleavage without dislocation activity) with low computational cost \citep[for details see][]{Brink2019, Brink:2022}.

We used the setups shown in \Cref{fig:setup}. For two simulations (a--b), we prepared a nanocrystalline material with grain size of $3\,\mathrm{nm}$, obtained by the Voronoi tesselation method \citep{Derlet2003}. For one simulation (c) we used a single-crystalline material with the (100) surface showing in $z$ direction. Initially, the top and bottom bulk regions (first bodies) each had a size of around $54\times54\times20\,\mathrm{nm^3}$ (total of around 8\,million atoms in the final simulation cell). For one nanocrystal and for the single crystal, we started from synthetic rough surfaces \citep{Wu2000} which were generated with a Hurst exponent of 0.8, a lower wavelength cutoff of $0.5\,\mathrm{nm}$, an upper wavelength cutoff of $27\,\mathrm{nm}$, no roll-off, and an RMS of heights of $2\,\mathrm{nm}$ using the software Tamaas \citep{tamaas}. The other nanocrystal had a flat surface initially. We then introduced four rigid particles each into the gaps between the surfaces (third bodies). These particles were polyhedra (rhombicuboctahedral shape), except for case (a), where we used round particles for comparison. Their diameter was chosen at $\approx 16\,\mathrm{nm}$. Note that the particles get coated quickly by material picked up from the surfaces. Therefore (i) their initial shape does not matter and (ii) the simulation resembles more closely the adhesive wear case than the abrasive one \citep{Brink:2022}.

The sliding simulations were performed with periodic boundary conditions along $x$ and $y$. A layer of thickness $0.4\,\mathrm{nm}$ was fixed at the ends of the top and bottom surfaces, where a normal force of $7.69\,\mathrm{\mu N}$ was applied, corresponding to an average pressure of approximately $2.6\,\mathrm{GPa}$ or 8\% of the hardness. Next to these boundary layers, another 0.4-nm-thick layer was used each to apply Langevin thermostats at room temperature with a damping constant of $0.01\,\mathrm{ps}$. The center-of-mass velocity of the layer was subtracted from the thermostating calculation to avoid an artificial drag force. Sliding was imposed on the top first body with a velocity of $20\,\mathrm{m/s}$ to a total sliding distance of $1\,\mathrm{\mu m}$ at an angle of $8.5^\circ$ off the x direction to avoid that the particles wear the same trench over and over again.

Every $0.1\,\mathrm{\mu m}$ sliding distance ($5\,\mathrm{ns}$), the simulation state was recorded. For post-processing, the third bodies are fixed in place while the surfaces are separated \citep{Brink:2022}. The atoms on the surfaces of the first bodies were indentified using a surface mesh generation algorithm \citep{Stukowski2014} implemented in Ovito \citep{Ovito}. This algorithm is based on testing if a virtual probe sphere of radius $0.385\,\mathrm{nm}$ can penetrate the material or not \citep{Stukowski2014}. The resulting surfaces are then analyzed as described in the following.

\subsection{Post-processing}
\label{Sec:methods_postproc}

\subsubsection{2D surfaces}

The height of a surface is defined by a function $h=h(x,y)$, where $x \in [-L_x/2,+L_x/2]$ and $y \in [-L_y/2,+L_y/2]$ span the surface. If we assume a periodic boundary conditions, it follows that $h(-L_x/2,y)=h(+L_x/2,y)$ and $h(x,-L_y/2)=h(x,+L_y/2)$. 

Periodicity allows analyzing the surface using Fourier series and, since we work with discrete datasets, discrete Fourier transform. 
The first step is to interpolate among the original irregular mesh points to then evaluate the interpolant on a $N \times N$ regular mesh, $N$ being the number of points in either direction. 
We use a piece-wise constant interpolant between atoms. 
See that by the end of this procedure one may obtain a set of points that do not reflect the periodicity of the original point cloud; this motivates the use of ``windowing'' discussed later. 

Heights are known in discrete fashion. The position of the $i$-th surface atom comes given by a triple $(x_i,y_i,z_i)$, $z_i = h(x_i,y_i)$ being the height. 
The set of all points forms an unstructured mesh which must be re-sampled into a regular grid (sampling intervals $\Delta x = \Delta y = 1 \, \text{\r{A}}$) before using \textit{discrete} Fourier transform methods. 
Then, the height of the topography features can be expressed as:

\begin{align} \label{eq:fourier_transform}
    h(x_i,y_i) 
    =
    {1 \over L_x L_y}
    \sum_{q_x,q_y}
    \hat{h}_{q_x,q_y}
    \exp{
    \left[
    \mathrm{i} 
    \left(
    {q_x x_i }
    +
    {q_y y_i}
    \right)
    \right]
    } \, ,
\end{align}
where the wavenumbers appear, horizontal $q_{x} = 2\pi n / L_x$ and vertical $q_{y} = 2\pi n/ L_{y}$, taking possible values indexed by $n \in [0,\ldots,N-1 ]$.
%
The amplitude corresponding to each combination of wavenumbers is computed as
\begin{align}
    \hat{h}_{q_x,q_y}
    =
    \sum_{x,y}
    h(x_i,y_i)
    \exp{
    \left[
    -\mathrm{i} 
    \left(
    {q_x x_i }
    +
    {q_y y_i}
    \right)
    \right]
    } \, .
\end{align} 
The heights are rescaled beforehand to guarantee $\sum_{i=1}^{N \times N} h(x_i,y_i) = 0$, what amounts to $\hat{h}_{0,0} = 0$.
%
%
The set of all Fourier coefficients will be referred henceforth as ``the spectrum of the surface'', and any individual coefficient as ``a harmonic''. 
%


The $h^2_{rms}$ is an important parameter when it comes to test if a surface is ``self-affine''. 
This parameter represents an average squared height \citep{Jacobs:2017}, and thus it conveys the magnitude of the topography oscillations. 
The 2D power spectral density (PSD) is a function of the wavenumbers defined using the harmonics' amplitudes 
\begin{align}
\label{eq:C2D}
    C^\text{2D}_{q_x,q_y}
    =
    {1 \over N^2}
    |\hat{h}_{q_x,q_y}|^2 
    \, ,
\end{align}
which is equivalent to the magnitude of the Fourier transform of the height-to-height autocorrelation function \citep{Jacobs:2017}, 
a consequence of Parseval's theorem \citep{Evans}. 
 
The lack of periodicity associated to discreteness and interpolation can introduce spurious high-frequency oscillations in the spectrum of the surfaces \citep{Jacobs:2017}. 
To avoid this issue, windowing is used. In this text, we use radial Hahn window, implicitly assuming that the roughness we are dealing with is isotropic. The radially-symmetric Hahn window is defined as \citep{Jacobs:2017}
\begin{align}
    w(x,y) 
    =
    \left( 
    {3 \pi \over 8}
    -
    {2 \over \pi}
    \right)^{-1/2}
    \left\{ 
    1 
    +
    \cos \left[
    {2 \pi \sqrt{(x-L_x/2)^2 + (y-L_y/2)^2}
    \over
    \mathrm{min}(L_x,L_y)
    }
    \right]
    \right\} \, ,
\end{align}
defined like this for $\sqrt{x^2 + y^2} < \mathrm{min}(L_x,L_y)/2$ and equal to zero everywhere else. The modified ``windowed'' heights are given by $h_{\text{Hahn}}(x_i,y_i) = w(x_i,y_i) h(x_i,y_i)$ for all $(x_i,y_i)$ such that $\sqrt{x_i^2 + y_i^2} < \mathrm{min}(L_x,L_y)/2$, and $h_{\text{Hahn}}(x_i,y_i) = 0$ otherwise.



The Fourier transform comes given in terms of the horizontal wavenumber $q_x$ and the vertical one $q_y$. 
If the surface is isotropic, then the coefficients of the Fourier series depend on the wavenumbers through $q_r = \sqrt{q_x^2 + q_y^2}$, meaning that the spectral amplitude $C^{2D}$ must possess axial symmetry with respect to the origin of the $q_x - q_y$ plane. 
Thus, for any fixed $q_r$, we can define the radial average of 2D PSDs (implicitly assuming isotropy) as 
\begin{align}
\label{eq:Ciso}
    C^\text{iso}(q_r)
    =
    {1 \over N_{\theta}}
    \sum_{\theta}
    C^\text{2D}(q_r) \, ,
\end{align}
where $N_{\theta}$ is a number of angular probes. We probe at $\theta  \in [0, 2\pi/100, \ldots, 198\pi/100 ]$, i.e., we average over $N_{\theta}=100$ points for every fixed $q_r$. 

The PSDs that we obtain seem to reasonably satisfy this assumption. 
Thus, the 2D spectrum indexed by $q_x$ and $q_y$ is converted into a 1D one that depends on $q_r$. 
If the surface is isotropic and self-affine, then the spectral amplitudes must scale as $\sim q^{-2(1+\mathrm{H})}$, where $\mathrm{H}$ is the Hurst exponent. 

The value of $\mathrm{H}$ is obtained through the slope of the line fitted using logarithmic scales, discarding the roll-off phase \citep{Jacobs:2017}. 

\subsubsection{1D line scans}

We have also performed 1D scans on the surfaces. 
%
Their theory is briefly introduced next, for further details see \cite{Enrico} and the appendix A of \cite{Jacobs:2017}.

In the 1D case, given a height 1D scan $h_\text{1D}(x)$ along the x-direction, the PSD (per unit length, at a wavelength $q_n$) of self-affine surfaces comes given as
\begin{align}
    {1 \over L_x}
    \left|
    \int_{-L_x/2}^{L_x/2}
    h_\text{1D}(x) 
    \exp{(-\mathrm{i} q_n x)}
    \right|^2
    dx
    \approx
    \Delta x
    {1 \over N}
    \left|
    \sum_{k=0}^{N-1}
    h_\text{1D}(x_k) 
    \exp{(-\mathrm{i} q_n x_k)}
    \right|^2
    \, ,
\end{align}
which has been discretized using a regular step $\Delta x = L_x / N$, thus $x_k \in [0, \Delta x, \ldots, (N-1)\Delta x]$. Therefore, the discrete spectrum of the PSD contains discrete wavenumbers $q_n = 2 \pi n / L_x $ for $n = 0, 1, 2, \ldots$ Thus
\begin{align}
    C^\text{1D} (q_n)
    =
    {1 \over N}
    \left|
    \sum_{k=0}^{N-1}
    h_\text{1D}(x_k) 
    \exp{(-\mathrm{i} q_n x_k)}
    \right|^2 \, .
\end{align}
    
which, after averaging over many scans \citep{Jacobs:2017}, must satisfy $ \sim q^{-(1+2 \mathrm{H})}$ if the roughness is self-affine. 

The height-to-height correlation function, defined as $\Delta h (\delta x) = \left< [ h(x + \delta x) - h(x) ]^2 \right>^{1/2}$ (where $\left< \cdot \right>$ means taking the spatial average), is another 1D statistical quantity of interest. 
It is known to follow $\Delta h (\delta x) \sim (\delta x)^\mathrm{H}$ if the 1D surface is indeed self-affine. 

Results of 1D analyses must be averaged across many scans to render the results consistent with the 2D results \citep{Jacobs:2017}. We use ten scans along x-direction and ten more along y.


\section{Results}
\label{Sec:results}

\subsection{Visualization of roughness evolution}

\begin{figure}
    \centering
    \includegraphics[width=0.99\linewidth]{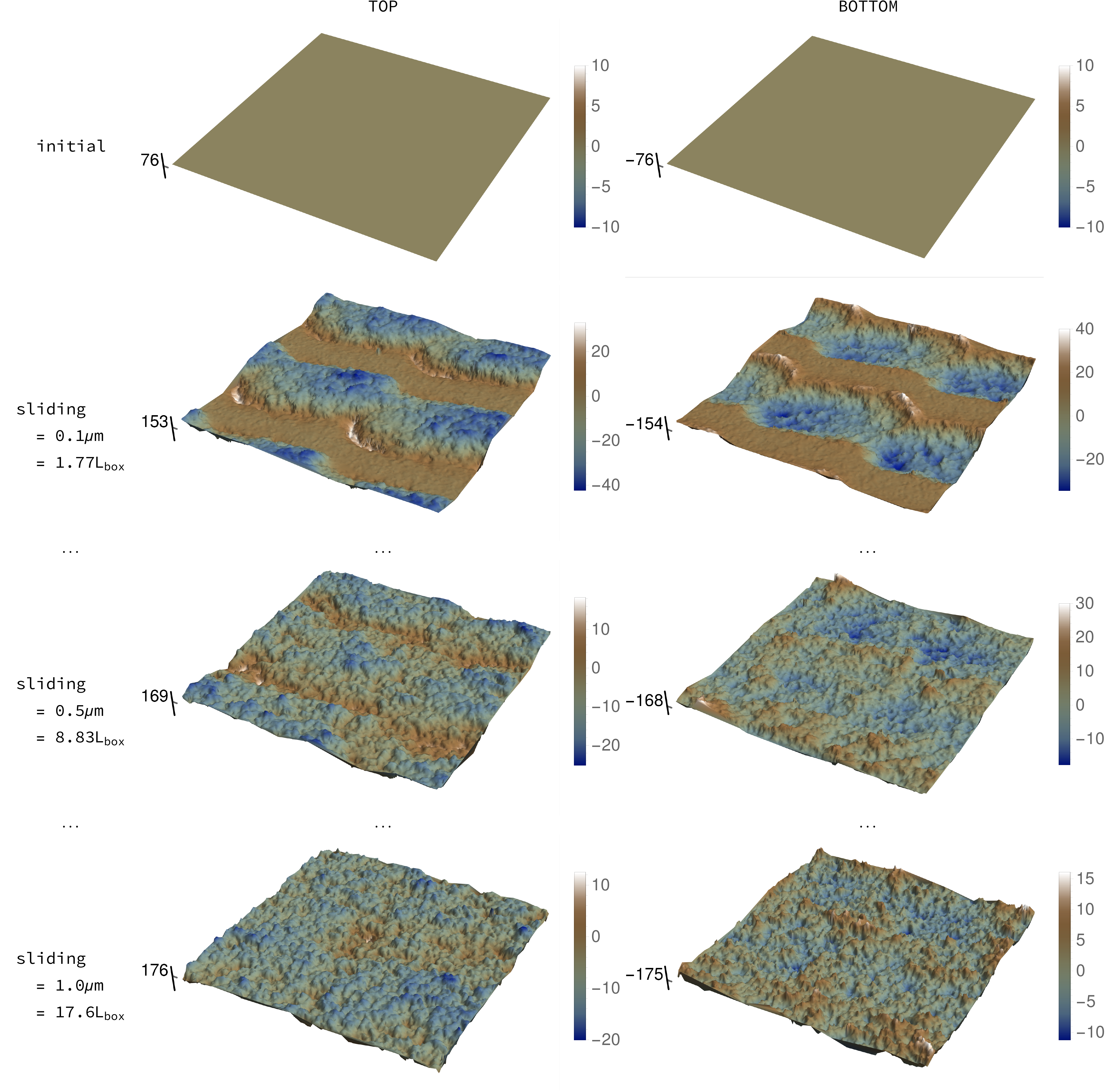}
    \caption{Snapshots of surface evolution (units in \r{A} unless otherwise stated): silicon surfaces (top and bottom) initially flat. Corresponding Hurst exponent evolution in \Cref{fig:hurst_Si_abrasive-on-flat-with-grains-and-round-particles}. Note changing scales. Vertical axes mark the (evolving) mean height of each surface. In either surfaces, white means topographical features ``bulging out'', while blue means penetrating into the surface bulk.}
    \label{fig:surfaces_AoF}
\end{figure}
\begin{figure}
    \centering
    \includegraphics[width=0.99\linewidth]{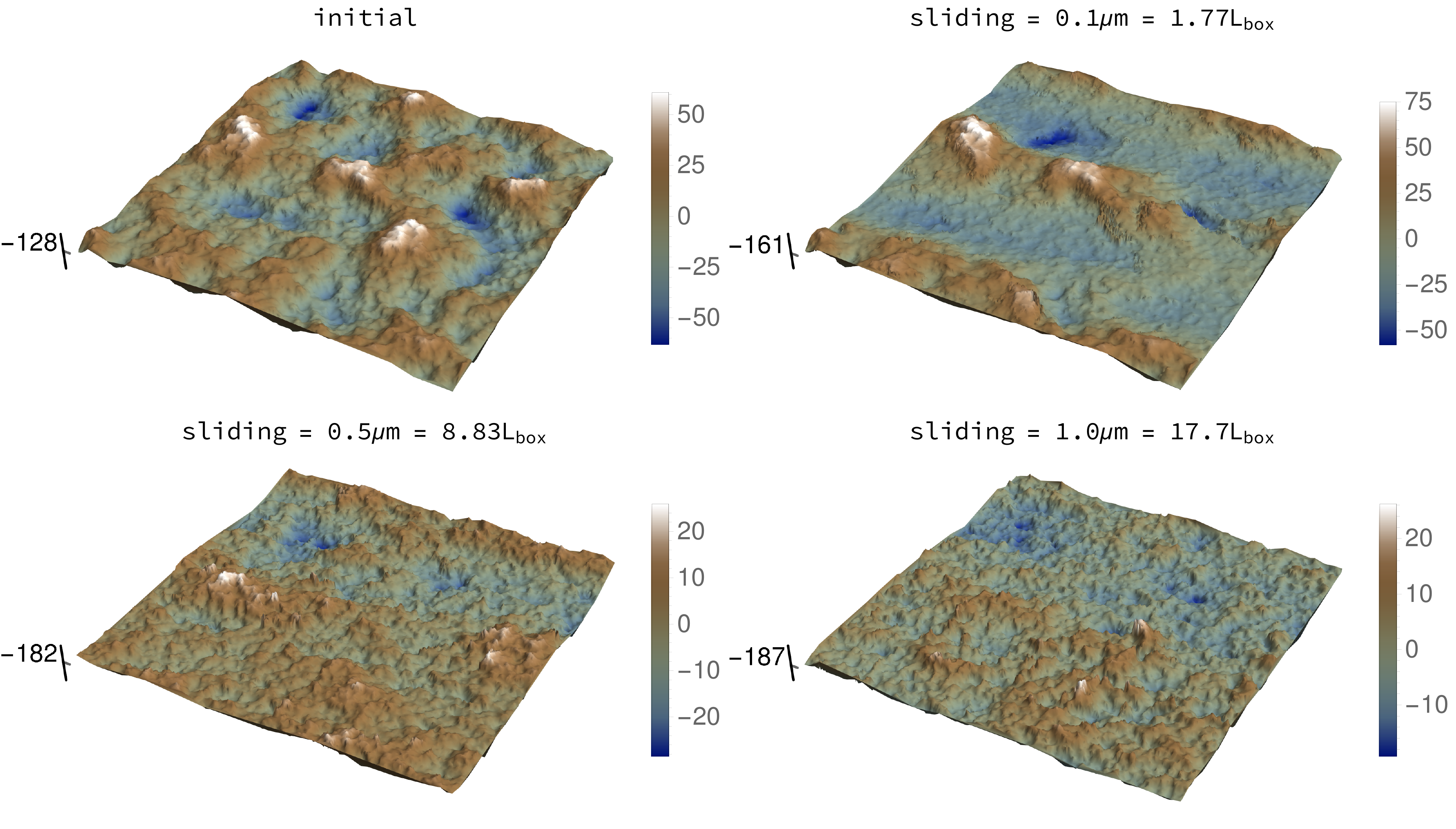}
    \caption{Snapshots of surface evolution (units in \r{A} unless otherwise stated): silicon surfaces (bottom) initially rough, nanocrystalline bulk. Corresponding Hurst exponent evolution in \Cref{fig:hurst_Si_abrasive-on-initially-rough-with-grains}. Note changing scales. Vertical axes mark the (evolving) mean height of each surface. In either surfaces, white means topographical features ``bulging out'', while blue means penetrating into the surface bulk.}
    \label{fig:surfaces_AoF_grains}
\end{figure}

For the geometrical setting and silicon-like material described in the \Cref{Sec:methods_tobias}, we show schemes of the roughness evolution in three different cases. 
The first, \Cref{fig:surfaces_AoF}, corresponds to the surfaces (top and bottom) that are initially flat, and whose bulk contains grain boundaries. 
The second one, \Cref{fig:surfaces_AoF_grains}, features an initially-rough isotropic surface (bottom one) created with Tamaas \citep{tamaas} with an initial Hurst exponent of $0.8$ whose bulk material also contains grain boundaries. 
Finally,
supplementary material Figure A.1.
,is similar to \Cref{fig:surfaces_AoF_grains}, but the bulk material is crystalline.

Every plot is accompanied by a scale to measure the amplitude of oscillations with respect to the mean. See that the color code remains the same, but the range of the scales changes between surfaces, since the magnitude of the topographical features evolves. 
The absolute position on the mean plane is marked in the vertical axis to better appreciate how the surface level descends as surface atoms are transferred to the coating of the third body. 
Note that the middle point between mean planes of the surfaces corresponds to the height equal to $0$. 
Lighter colors highlight features that ``stick out'' of the surface, while darker ones penetrate into the bulk.

``Trenches'' associated to the scratching by the third body are observed in both top and bottom surfaces after sliding by $0.1 \, \mu \mathrm{m}$ (see second row in \Cref{fig:surfaces_AoF}). 
Bear in mind that the particles' trajectories wind over the whole surface, owing to the sliding direction being not aligned with either axis of the surface.  
As sliding progresses, the topography ``homogenizes'': the initially-flat surfaces become isotropically rough (third and fourth row in \Cref{fig:surfaces_AoF}) and the initially-rough ones evolve to a new state, similarly isotropic but characterized by lower height amplitudes 
(e.g., \Cref{fig:surfaces_AoF_grains}).

\subsection{2D PSDs}

We also provide graphics of PSD evolution 
in the supplementary material.
%
They clearly reveal ``textured'' surfaces during the first sliding stages, meaning that we see light spots in figures in supplementary material section C, 
indicating large magnitude oscillations that break the radial symmetry required by isotropy. 
Physically, this marked wavelengths are related to the deep grooves left by the wear particles during the first runs, which scratch the initial surface in an abrasive manner. 
As more sliding unfolds, the debris is coated with surface atoms, changing the wear mechanism to tearing of small flakes of material, induced by adhesion and the debris' rolling movement \citep{Milanese:2020b}. 
The texture then fades away, yielding a PSD wavenumber distribution that seems reasonably angle independent, i.e., isotropic, see the final sliding PSD figures in supplementary material section C. 
%
This result is reassuring insofar it backs our subsequent analysis as to self-affinity. 

\subsection{Logarithmic slope of radial PSD (\textit{C}\textsuperscript{iso}): fitting the Hurst exponent}

As explained in the Methods section, isotropic surfaces depend solely on the modulus of the wavenumber vector $(q_x,q_y)$, not on the ratio $q_x/q_y$. 
This motivates the definition of a radial PSD, characterized by $C^\text{iso}$, \cref{eq:Ciso}. 
The Hurst exponent can be easily fitted when this function is expressed in log--log scales. This exercise has been pursued in 
supplemenrary material section B. 
%

\subsection{Hurst exponent evolution}

\begin{figure}
\centering
\captionsetup[subfigure]{justification=centering}
\begin{subfigure}[b]{.45\linewidth}
\includegraphics[width=\linewidth]{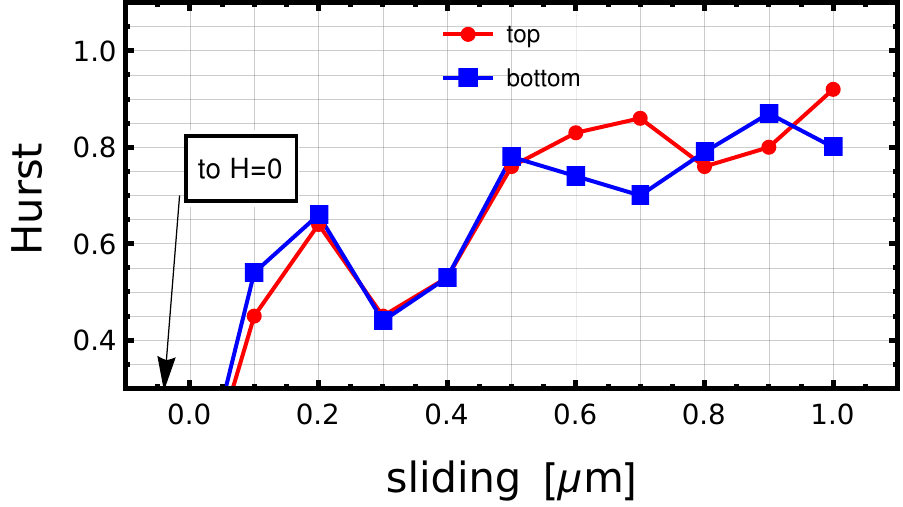}
\caption{ (a) Initially flat, bulk material nanocrystalline.}    \label{fig:hurst_Si_abrasive-on-flat-with-grains-and-round-particles}
\end{subfigure}
\begin{subfigure}[b]{.45\linewidth}
   \includegraphics[width=\linewidth]{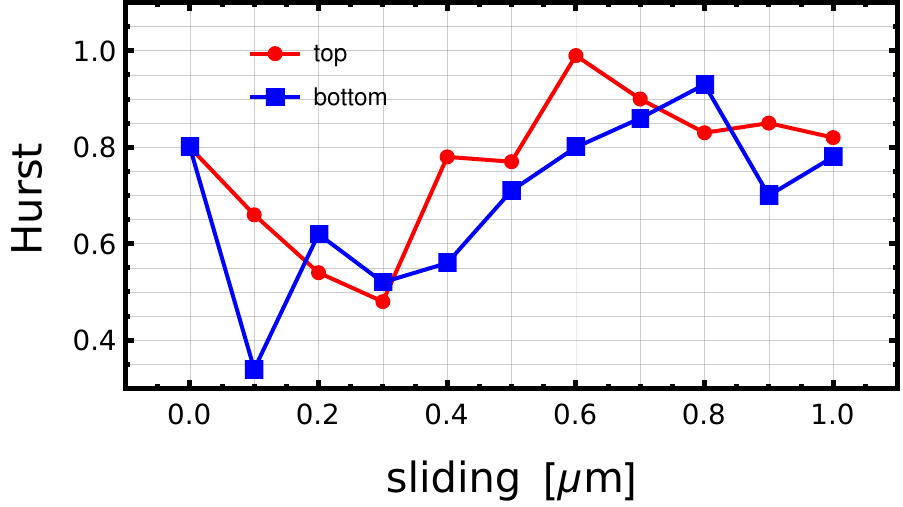}
    \caption{ (b) Initially rough ($\mathrm{H}=0.8$), bulk material nanocrystalline.}
    \label{fig:hurst_Si_abrasive-on-initially-rough-with-grains}
\end{subfigure}
\begin{subfigure}[b]{.45\linewidth}
   \includegraphics[width=\linewidth]{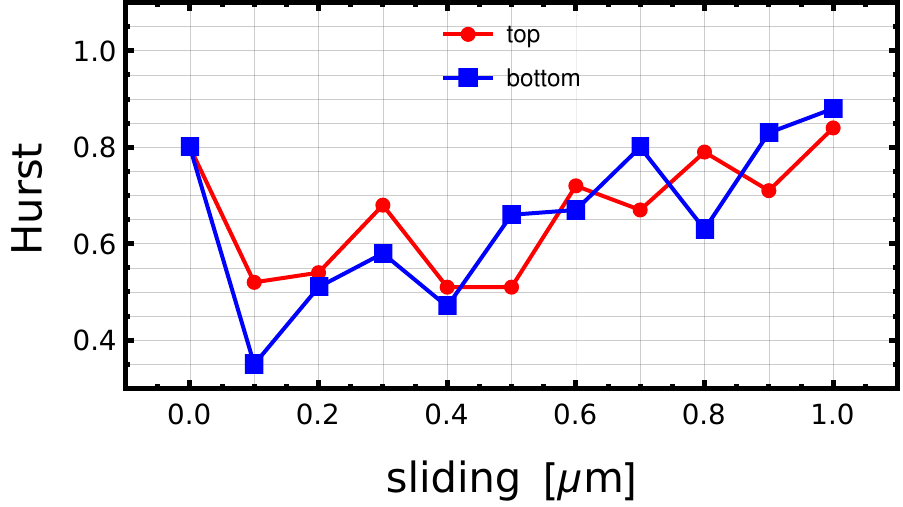}
    \caption{ (c) Initially rough ($\mathrm{H}=0.8$), bulk material single crystal.}
    \label{fig:hurst_Si_abrasive-on-initially-rough-single-crystal}
\end{subfigure}
\caption{Evolution of the Hurst exponent of Si surfaces worn by wear particles during sliding: three configurations.}
\end{figure}

\Cref{fig:hurst_Si_abrasive-on-flat-with-grains-and-round-particles,fig:hurst_Si_abrasive-on-initially-rough-with-grains,fig:hurst_Si_abrasive-on-initially-rough-single-crystal} show the roughness evolution corresponding to the cases presented in
Figures 2, 3 and A.1. (supplementary material),
respectively. The method to extract the exponent from the discrete data was outlined in \Cref{Sec:methods_postproc}. The initial values (before any sliding) are found to be, as expected, $0$ in the case of initially-flat surface and approximately $0.8$ in the two initially-rough cases.

\begin{center}
\begin{tabular}{c| c c } 
 Surface & 1D analysis & 2D analysis \\ 
 \hline
     Initially flat, nanocrystalline bulk & 0.76 & 0.86  \\
    Initially rough, nanocrystalline bulk & 0.81 & 0.80  \\
    Initially rough, crystalline bulk & 0.78 & 0.86  \\
 \hline
\end{tabular}
\captionof{table}{Comparison final Hurst exponents, derived from two different methods. 2D-analysis results represent top-and-bottom average, while 1D ones correspond to averages over top and bottom surfaces, and twenty line scans each, ten along the x direction and ten more along y.}
\label{table:comparison}
\end{center}

\subsection{1D line scans}

Since the range of data for the PSDs does not span many decades, it is necessary to verify the self-affinity using further measures in addition to $C^\text{iso}$.
Averaged 1D line scans of the final surfaces match reasonably well with the ones obtained directly from processing the complete 2D surfaces. 
See \Cref{table:comparison} for average Hurst exponents from 2D analysis (average between the values inferred from the top and bottom surfaces) and the 1D analysis (averaged over twenty scans, ten along the x direction and ten along y).


We provide plots of such results in 
supplementary material. 
%
Moreover, these scans yield an important piece of information: the Hurst exponents obtained from the 1D PSD predict well the ones of the height-to-height correlation function. 
This provides a consistency in self-affinity that palliates the lack of roughness information over many decades \citep{Enrico}, as would be desirable. 
%
Supplementary material figures,  section B,
show the fitting of the Hurst exponent from the 1D PSD on the left panel, while the right one shows the height correlation and the self-affine slope \textit{presumed} from the Hurst exponent obtained from the PSD. We acknowledge that the exponent obtained from PSD analysis matches well the slope in the short correlation lengths, before the curves level. 
These plateaus already occur for height differences above the order of $1\,\mathrm{nm}$, but this is simply a result of the low $h_{\text{rms}}$ values, which in turn are caused by the depletion of the bigger topography features during the wear process: correlations are only possible for features that actually exist, i.e., up to the order of $h_{\text{rms}}$.

\section{Discussion}
\label{Sec:discussion}

The main insight we can extract, c.f. \cite{Enrico}, it is that, for this material and system size and configuration, the topography of the three surfaces seems to converge to a steady state, characterized by $\mathrm{H} \approx 0.8$, independently both of the initial conditions (rough or flat) and  of the microstructure of the bulk material. 
The latter remark is in agreement with the findings of \cite{Hinkle:2020}: material heterogeneity can have a strong influence on wear at this scale \citep{Wattel:2022}, but it cannot be the controlling factor of roughness evolution; rather, discrete deformation mechanisms  bear responsibility \citep{Irani:2019,Hinkle:2020}; in our case, they amount primarily to wear by tearing of flakes of material \citep{Milanese:2020,Milanese:2020b}. 
%

It is also remarkable how the initially-flat surface, \Cref{fig:surfaces_AoF,fig:hurst_Si_abrasive-on-flat-with-grains-and-round-particles} appears to converge faster to the steady-state roughness regime, with smaller oscillations around a mean slightly greater than 0.8. 
This seems to indicate that the ``memory'' of the previous roughness in the other two cases takes longer to be erased.
This is reminiscent of the memory length-scale that appears in rate-and-state friction laws \citep{Scholz:2019}. 
In this context, the frictional state of the interface changes dynamically, reaching the new one after a transient. The extent of this transient is considered a function of the existing microcontacts, i.e, of the roughness. 
The numerical results seem to reflect this, since the ``microcontact population'' of the flat surface is very different from the one of the intially-rough surface.
Micromechanically, this could be related to the need of attaining an intermediate ``indifferent'' roughness state (corresponding to $ \mathrm{H} \approx 0.5$, in which the probability of the vertical position of the next atoms is equally probable to be below or above the current one, i.e., the roughness follows a standard random walk). 
A similar observation is made by \cite{Hinkle:2020}. 
From the prior samples, it seems that the roughness may need to oscillate around these values before reaching the steady-state. In order to evolve into the intermediate regime, the features associated to the roughness  $\mathrm{H} \approx 0.8$ have to be \textit{erased} by wear, while in the flat case they are directly \textit{created} by wear.

As mentioned in the introduction, simulations involving more ductile materials (either zinc, copper or aluminum) converged to surface welding in which both surfaces were joined together, plastically deforming to engulf the particles in their midst. 
For this brittle materials, the destiny may be the same in the long term: we acknowledge an ever-growing volume of the coated third bodies \citep{Brink:2022}, which could ultimately agglomerate and bridge the gap between surfaces forming the aforementioned shear-band state. Hence, this steady-state could be conceived as an ``intermediate asymptotics'' state, that may seem locally stable but that can devolve into a shearband-like state over longer timescales. 
In practice, this final state may be avoided by other mechanisms acting on the said extended time spans. 
One of such is passivation: the coating atoms may react with the atmosphere and form in turn a composite layer that prevents the flake tearing process, thus deactivating the welding.



\section{Final remarks}
\label{Sec:final_remarks}


The roughness evolution induced by third-body wear has been studied using large-scale molecular dynamics simulations. 
Using conventional post-processing techniques \citep{Jacobs:2017}, the self-affinity of the resulting surfaces has been verified, yielding in addition evidence as to the existence of convergence to roughness with $\mathrm{H} \approx 0.8$. 
This final state, induced under certain circumstances (e.g., brittle enough material and abrasive wear particles), appears to display similar roughness characteristics independently of both the initial topography (either flat or rough with a higher surface roughness $\mathrm{H} \approx 0.8$) and bulk lattice structure (either single crystal or presence of grain boundaries). 

We stress that this independence from bulk structure is consistent with \cite{Hinkle:2020}, and that these authors also reported the existence of a transient ``random-walk'' state ($\mathrm{H} \approx 0.5$) during the evolution of roughness prior to attainment of a steady state. 
We also emphasize that \cite{Hinkle:2020} run purely plasticity-driven simulations, wherein neither third bodies nor fracture were present.




\section*{Acknowledgments}
J. G.-S. and J.-F. M. gratefully acknowledge the sponsorship of the Swiss National Science Foundation (grant \#197152, ``Wear across scales''). 
Computing time was provided by a grant
from the Swiss National Supercomputing Center (CSCS) under project IDs s784 (“The evolution of rough surfaces in the adhesive wear regime”) and s972 (“Surface and subsurface evolution of metals in three-body wear conditions”), as well as by École polytechnique fédérale de
Lausanne (EPFL) through the use of the facilities of its Scientific IT and Application Support Center.

\section*{Supplementary material}

Details concerning the intermediate results to obtain Hurst exponents are contained in the supplementary material document. 
The repository titled \texttt{roughness\_wear} within the first author Github page (\texttt{github.com/jgarciasuarez}) contains Mathematica notebooks \citep{Mathematica} detailing data post-processing leading to the results displayed in the text.

\bibliography{references}

\end{document}